# Superconductivity in alkaline earth metal–filled skutterudites Ba$_x$Ir$_4$X$_{12}$ (X = As, P)

Yanpeng Qi[†*], Hechang Lei[†], Jiangang Guo[†], Wujun Shi[§∥], Binghai Yan[#], Claudia Felser[§] and Hideo Hosono[†‡*]

[†]Materials Research Center for Element Strategy, Tokyo Institute of Technology, 4259 Yokohama, Japan

[‡]Laboratory for Materials and Structures, Tokyo Institute of Technology, 4259 Yokohama, Japan

[§]Max Planck Institute for Chemical Physics of Solids, 01187 Dresden, Germany

[∥]School of Physical Science and Technology, ShanghaiTech University, 200031 Shanghai, China

[#]Department of Condensed Matter Physics, Weizmann Institute of Science, Rehovot 7610001, Israel

*Supporting Information Placeholder*

**ABSTRACT:** We report superconductive iridium pnictides Ba$_x$Ir$_4$X$_{12}$ (X = As and P) with a filled skutterudite structure, demonstrating that Ba filling dramatically alters their electronic properties and induces a nonmetal-to-metal transition with increasing the Ba content $x$. The highest superconducting transition temperatures are 4.8 and 5.6 K observed for Ba$_x$Ir$_4$As$_{12}$ and Ba$_x$Ir$_4$P$_{12}$, respectively. The superconductivity in Ba$_x$Ir$_4$X$_{12}$ can classified into the Bardeen–Cooper–Schrieffer type with intermediate coupling.

Caged compounds exhibit crystal structures composed of rigid covalently bonded cage-forming frameworks enclosing differently bonded guest atoms, being not only of scientific but also of significant technological interest. The ability of these materials to accommodate guest filler species results in a wide range of attractive features[1-4] such as heavy-fermion state formation,[5,6] multipole magnetic ordering,[7,8] Fermi- and non-Fermi liquid behavior,[9] good thermoelectric properties,[3,10] and conventional or unconventional superconductivity.[6,11-17]

Among caged compounds, much attention is directed at investigating filled skutterudites, which exhibit a general formula of AT$_4$X$_{12}$, where A is an electropositive cation, T is a transition metal, and X is a pnicogen (P, As, or Sb). The skutterudite framework features eight tilted octahedra per unit cell enclosing two icosahedral cages, with a transition metal atom located at the center of each octahedron. This simple structure acts as a prototype for a large class of compounds, with its flexibility allowing the incorporation of different guest atoms, from alkali and alkali earth to rare earth metals. Although iridium skutterudites (IrAs$_3$ and IrP$_3$) have already been known for a long time,[18,19] no filled versions have been reported so far.

The discovery of high-temperature superconductivity in iron pnictides[20] has rekindled the exploration of novel metal pnictide- and chalcogenide-based superconductors.[21,22] Herein, we report the high-pressure synthesis of Ba-filled skutterudites Ba$_x$Ir$_4$X$_{12}$ (X = As, P), showing that these compounds exhibit bulk superconductivity with transition temperatures of 4.8 K (Ba$_{0.85}$Ir$_4$As$_{12}$) and 5.6 K (Ba$_{0.89}$Ir$_4$P$_{12}$). To the best of our knowledge, this is the first report of superconducting iridium pnictides with a filled skutterudite structure. The basic structural features of Ba$_x$Ir$_4$X$_{12}$ were examined by powder X-ray diffraction (XRD) and electron probe microanalysis (EPMA), and parameters of normal and superconducting states were also established.

Samples with a nominal composition of Ba$_x$Ir$_4$As$_{12}$ and Ba$_x$Ir$_4$P$_{12}$ were synthesized by high-pressure solid-state reactions starting from BaAs/BaP, Ir, and As/P. BaAs and BaP were synthesized by heating a stoichiometric mixture of Ba and As or P at 973 K for 10 h. The starting material mixture of the desired composition was placed in an $h$-BN capsule and heated at 1473–1573 K and 5 GPa for 4 h using a belt-type high-pressure apparatus. All starting materials and precursors were prepared in a glove box filled with purified Ar gas (H$_2$O, O$_2$ < 1 ppm).

The crystalline phases of the thus prepared samples were identified by powder XRD (D8 ADVANCE (Cu rotating anode), Bruker). Rietveld refinement of XRD patterns was performed using TOPAS3 software. Elemental compositions were determined using EPMA (JXA-8530F, JEOL). The dependence of direct-current electrical resistivity ($\rho$) on temperature was measured at 2–300 K by a conventional four-probe method. Magnetization ($M$) measurements were performed using a vibrating sample magnetometer (Quantum Design MPMS). Specific heat data were obtained by a conventional thermal relaxation method (Quantum Design PPMS).

Density functional theory (DFT) calculations were performed using the Vienna *Ab initio* Simulation Package (VASP)[23] with a plane-wave basis, with interactions between valence electrons and ion cores described using the projector-augmented wave method.[24,25] Exchange and correlation energies were formulated using the generalized gradient approximation with the Perdew–Burke–Ernzerhof functional.[26] An experimentally determined lattice constant was used, and atom positions were relaxed until the forces acting on the atoms were less than 5 meV Å$^{-1}$. Spin-orbit coupling was included in the electronic structure calculations.

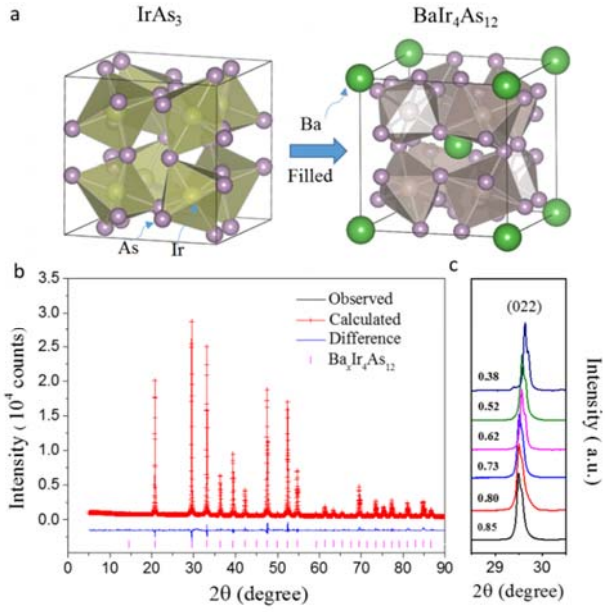

Figure 1. Crystal structure and powder X-ray diffraction pattern of $Ba_xIr_4As_{12}$. (a) Crystal structures of $IrAs_3$ and $Ba_xIr_4As_{12}$. (b) Observed and fitted XRD patterns of $Ba_xIr_4As_{12}$. (c) Position of the (022) XRD peak as a function of $x$.

Figure 1a shows the crystal structures of $IrAs_3$ and Ba-filled $Ba_xIr_4As_{12}$, with the latter featuring Ba-filled icosahedra of As atoms. The Ba content of $Ba_xIr_4As_{12}$ was determined by EPMA as an average of eight measurements (Figure S1 and Table S1), linearly increasing with increasing $x$, with saturation occurring at $x \geq 0.85$ for $Ba_xIr_4As_{12}$ and at $x \geq 0.89$ for $Ba_xIr_4P_{12}$ (Figure S2). Figure 1b shows the XRD pattern and Rietveld refinement result for $Ba_xIr_4As_{12}$. Except for minor impurity peaks attributed to $IrAs_3$, all other peaks could be attributed to a filled skutterudite structure (cubic, space group $Im\bar{3}$ (No. 204), $Z = 2$), shifting to smaller angles with increasing Ba content due to successful filling (Figure 1c and Figure S3). The lattice parameter ($a$) linearly increased from 8.4697 Å at $x = 0$ to 8.5605 Å at $x = 0.85$, with a similar phenomenon also observed for $Ba_xIr_4P_{12}$ (Figure S4).

Figure 2a displays the temperature dependence of the electrical resistivity of $Ba_xIr_4As_{12}$ between 2 and 300 K. The corresponding data for $IrAs_3$ indicate semiconducting behavior, in agreement with the calculated band structure. However, Ba filling results in dramatically altered electronic properties: whereas a moderate change is observed at low $x$ ($x < 0.15$), a further increase of Ba content results in a sudden increase of resistivity, which reaches a maximum at $x = 0.52$, with a rapid resistivity decrease observed as $x$ is increased further. The resistivity of $Ba_xIr_4As_{12}$ exhibits a non-monotonic evolution with increasing $x$, which is sharply different from other filled skutterudites.[1,3] This anomalous behavior is attributed to the change of charge carrier type, as revealed by the Hall resistivity ($\rho_{yx}$) sign change from positive to negative (indicating a change from $p$- to $n$-type conduction) with increasing $x$. In contrast to $Ba_xIr_4As_{12}$, no such change was observed for $Ba_xIr_4P_{12}$, which showed a gradually decreasing resistivity with increasing $x$ (Figures S5 and S6).

As $x$ was increased further, a sharp drop of $\rho(T)$ at a critical temperature $T_c = 3.2$ K was observed for $Ba_xIr_4As_{12}$, marking the onset of a superconducting transition. The above temperature increased with increasing $x$ to $T_c = 4.8$ K (Figure 2b). Figure 2d shows the resistivity of $Ba_xIr_4As_{12}$ under different magnetic fields, revealing that the superconducting transition was suppressed. The upper critical field ($\mu_0H_{c2}(0)$) was estimated as 6.9 T (Figure S7), yielding a Ginzburg-Landau coherence length $\xi_{GL}(0)$ of 6.9 nm. A strong diamagnetic signal was clearly observed below $T_c$ (Figure S8), indicating the bulk superconductivity of $Ba_xIr_4As_{12}$, which was further confirmed by the large specific heat jump at $T_c$ (Figures 3a, b). Therefore, the temperature dependence of $C_p$ was fitted as $C_p(T)/T = \gamma + \beta T^2 + \delta T^4$, which yielded an electronic specific heat coefficient ($\gamma_N$) of 35.0 mJ mol$^{-1}$ K$^{-2}$ and $\beta = 1.1$ mJ mol$^{-1}$ K$^{-4}$. The Debye temperature $\Theta_D$ equaled 311 K, and the normalized specific heat jump value, $\Delta C/\gamma_N T_c$, was estimated as 0.82, being substantially smaller than the Bardeen–Cooper–Schrieffer (BCS) weak coupling limit value of 1.43.[27] Using the McMillan formula for electron-phonon-mediated superconductivity,[28] we obtained the electron-phonon coupling constant of $Ba_{0.85}Ir_4As_{12}$ as $\lambda_{e-ph} = 0.63$, corresponding to a superconductor with intermediate coupling. The superconducting transitions of $Ba_xIr_4P_{12}$ are also illustrated in Figures 2c, e and Figure S9, with relevant parameters summarized in Table 1.

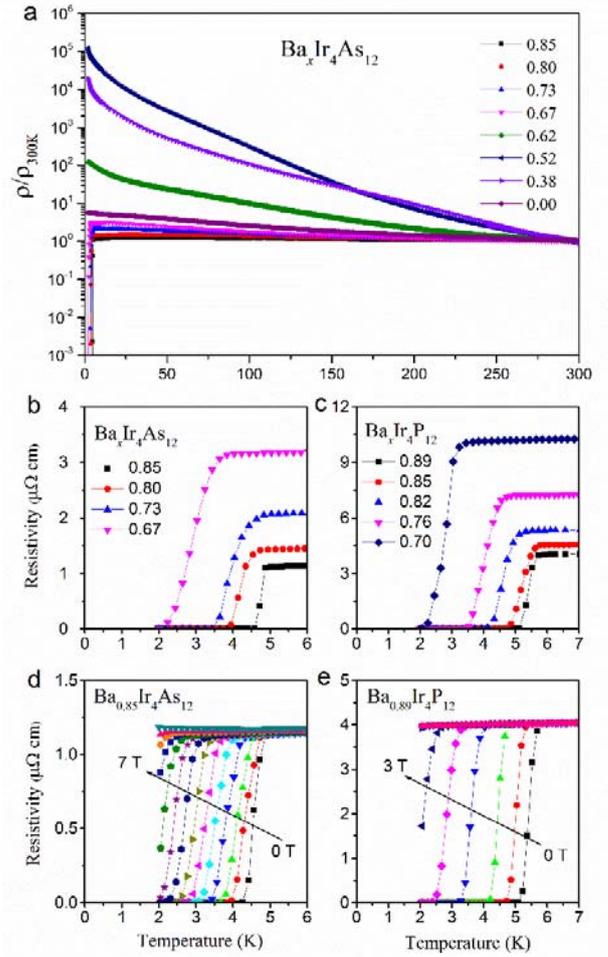

Figure 2. Electrical resistivity of $Ba_xIr_4As_{12}$ and $Ba_xIr_4P_{12}$ as a function of Ba content. (a) Electrical resistivity of $Ba_xIr_4As_{12}$ as a function of temperature. Electrical resistivity drop and low-temperature superconducting behavior of (b) $Ba_xIr_4As_{12}$ and (c) $Ba_xIr_4P_{12}$. Temperature dependence of resistivity under different magnetic fields for (d) $Ba_{0.85}Ir_4As_{12}$ and (e) $Ba_{0.89}Ir_4P_{12}$.

**Table 1. Structural and superconductivity-related parameters of $Ba_{0.85}Ir_4As_{12}$ and $Ba_{0.89}Ir_4P_{12}$.**

| Compound | $Ba_{0.85}Ir_4As_{12}$ | $Ba_{0.89}Ir_4P_{12}$ |
|---|---|---|
| Space group ($Z$) | Cubic $Im\bar{3}$- (No. 204), $Z = 2$ | |
| Lattice parameter $a$/Å | 8.5605 | 8.1071 |
| Transition temperature $T_c$/K | 4.8 | 5.6 |
| Upper critical field $H_{c2}$/T | 6.9 | 2.9 |
| Coherence length $\xi_{GL}(0)$/nm | 6.9 | 10.7 |
| Debye temperature $\Theta_D$/K | 311 | 368 |
| Einstein temperature $\Theta_E$/K | 92 | 150 |
| $\Delta C/\gamma T_c$ | 0.82 | 1.04 |
| Electron-phonon coupling constant $\lambda_{e\text{-}ph}$ | 0.63 | 0.62 |

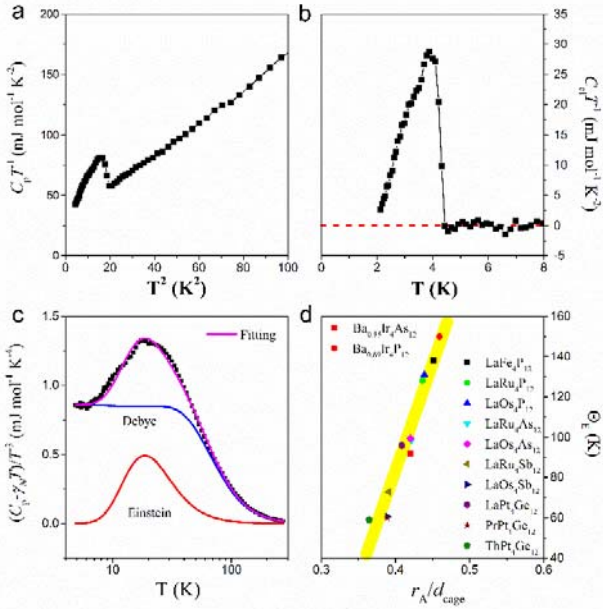

Figure 3. Specific heat and Einstein temperature of $Ba_{0.85}Ir_4As_{12}$. (a, b) Specific heat data for $Ba_{0.85}Ir_4As_{12}$ in the vicinity of the superconducting transition. (c) Normal-state specific heat data for $Ba_{0.85}Ir_4As_{12}$. Solid lines denote fitting curves discussed in the text, with blue and red lines representing the individual contributions of Debye and Einstein parts. (d) Plot of Einstein temperature vs. $r_A/d_{cage}$ for superconducting filled skutterudites.

Similarly to other caged compounds such as Si-Ge clathrates[29] and β-pyrochlore oxides,[30] an important structural feature of Ba-filled skutterudites is Ba atom rattling. Figure 3c shows the normal-state specific heat data for $Ba_{0.85}Ir_4As_{12}$, revealing broad peaks at ~20 K that manifest the presence of a low-energy Einstein vibration.[31-33] This so-called boson peak always involves local vibration modes[34] or transverse acoustic phonon anomalies.[35] Data fitting using a combination of Debye and Einstein oscillator models yielded Einstein temperatures $\Theta_E$ = 92 and 150 K for $Ba_{0.85}Ir_4As_{12}$ and $Ba_{0.89}Ir_4P_{12}$, respectively.

We subsequently considered characteristic energies of rattling, i.e., 92 K for Ba atoms in $Ba_xIr_4As_{12}$. The rattling energy has been sorted out in terms of the guest free space (gfs) in caged compounds.[36,37] Figure 3d shows plots of $\Theta_E$ vs. $r_A/d_{cage}$ for $Ba_xIr_4As_{12}$ and other superconducting filled skutterudites, where $r_A$ is the effective ionic radius for a 12-coordinated site, and $d_{cage}$ is the distance from the central atom A to the nearest cage-forming atoms (Table S2). The linear relationship between $\Theta_E$ and $r_A/d_{cage}$ strongly demonstrates that gfs is negatively correlated with rattling energy. Although low-energy phonons originating from the presence of guest Ba atoms may be crucial for the superconductivity of $Ba_xIr_4X_{12}$ (X = As, P), no simple relationship between $T_c$ and $\Theta_E$ analogous to that of β-pyrochlore oxides exists for filled skutterudites, implying a different superconductivity mechanism or the prevalence of other factors.[38,39]

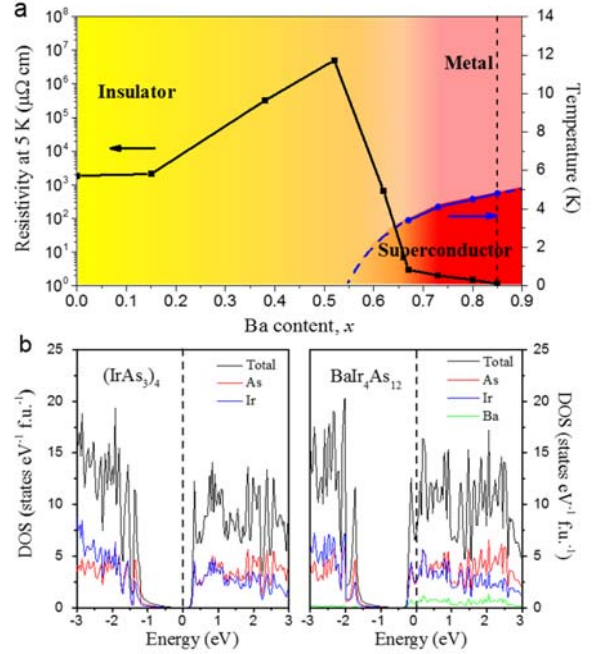

Figure 4. Phase diagram and electronic structures of $Ba_xIr_4As_{12}$. (a) Phase diagram of $Ba_xIr_4As_{12}$. (b) DOS for $(IrAs_3)_4$ and $BaIr_4As_{12}$ around the Fermi level, with the Fermi energy taken as zero.

Figure 4a shows the electronic phase diagram of $Ba_xIr_4As_{12}$, demonstrating that Ba filling dramatically alters the electronic properties of this compound. For $x < 0.15$, a weak enhancement of the overall magnitude of $\rho$ is observed. Subsequently, the value of $\rho(5\text{ K})$, i.e., normal-state resistivity just above the superconducting transition temperature, rapidly increases with increasing $x$, reaching a maximum at $x = 0.52$. For $x \geq 0.52$, a dramatic decrease of resistivity by more than six orders of magnitudes is observed, indicating a nonmetal-to-metal transition. Superconductivity starts to appear at $x \geq 0.67$, with $T_c$ monotonically increasing with $x$ and reaching a maximum of 4.8 K at $x = 0.85$ (samples with $x \geq 0.85$ could not be synthesized under the present conditions). A phase diagram was also obtained for $Ba_xIr_4P_{12}$, with a maximum $T_c$ of 5.6 K observed for $x = 0.89$ (Figure S10). The electronic band structure of $Ba_xIr_4As_{12}$ was investigated by DFT calculations (Figures S11 and S12). Figure 4b shows the calculated total and partial near-$E_F$ density of states (DOS) curves for $(IrAs_3)_4$ and $BaIr_4As_{12}$, revealing that $IrAs_3$ is a normal semiconductor with a narrow band gap ($\Delta E = 0.16$ eV), in agreement with the abovementioned experimental results. Ba filling did not significantly alter states around the Fermi level, which still mainly originated from the interactions of strongly hybridized As 4$p$ and Ir 5$d$ orbitals, whereas the Fermi level was shifted toward the conduction band (CB), finally crossing it. Ba 6$s$ states were mainly above the Fermi level. Overall, these results were consistent with Ba $6s$ to [Ir$_4$As$_{12}$] polyanion charge transfer.[16,17] For BaIr$_4$As$_{12}$, the total DOS at $E_F$ equaled 9.11 states eV$^{-1}$ f.u.$^{-1}$, corresponding to a coupling constant $\lambda = \gamma_N/\gamma_0 - 1 = 1.1$ and providing further evidence of intermediate coupling in BaIr$_4$As$_{12}$.

In conclusion, we identified Ba$_x$Ir$_4$As$_{12}$ and Ba$_x$Ir$_4$P$_{12}$ as new superconductors with a filled skutterudite structure, observing insulator-to-metal and metal-to-superconductor transitions with increasing Ba content. Analysis of normal-state specific heat data revealed the presence of a local low-energy Einstein mode originating from the presence of guest Ba atoms, which may favor superconductivity. Both experimental and DFT calculation results suggested that Ba$_x$Ir$_4$X$_{12}$ (X = As, P) can be categorized as Bardeen–Cooper–Schrieffer superconductors with intermediate coupling.

## ASSOCIATED CONTENT

### Supporting Information

The Supporting Information is available free of charge on the ACS Publications website.

Chemical composition and structure analysis of all samples, properties of other superconducting filled skutterudites, Hall Effect and superconducting properties, and DFT calculation results (PDF)

## AUTHOR INFORMATION

### Corresponding Author

E-mail: hosono@msl.titech.ac.jp

qi-yanpeng@hotmail.com

### Notes

The authors declare no competing financial interests.

## ACKNOWLEDGMENT

This work was supported by the Funding Program for World-Leading Innovative R&D on Science and Technology (FIRST) and the Element Strategy Initiative to form a Research Core by MEXT, Japan.

Authors are required to submit a graphic entry for the Table of Contents (TOC) that, in conjunction with the manuscript title, should give the reader a representative idea of one of the following: A key structure, reaction, equation, concept, or theorem, etc., that is discussed in the manuscript. Consult the journal's Instructions for Authors for TOC graphic specifications.

Insert Table of Contents artwork here

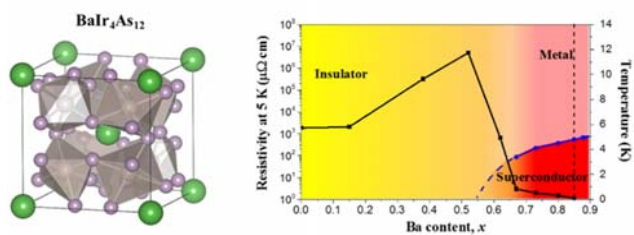